\documentclass[lettersize,journal]{IEEEtran}
\usepackage{amsmath,amsfonts}
\usepackage{amssymb}
\usepackage{algorithmic}
\usepackage{algorithm}
\usepackage{array}
\usepackage[caption=false,font=normalsize,labelfont=sf,textfont=sf]{subfig}
\usepackage{textcomp}
\usepackage{stfloats}
\usepackage{url}
\usepackage{verbatim}
\usepackage{graphicx}
\usepackage{comment}

\usepackage[backend = biber, isbn=false, style=ieee]{biblatex}
\renewbibmacro*{url+urldate}{}
\renewbibmacro*{doi+eprint+url}{}
\renewbibmacro*{isbn}{}

\usepackage{tikz}

\newcommand\submittedtext{%
  \footnotesize This work has been submitted to the IEEE for possible publication. Copyright may be transferred without notice, after which this version may no longer be accessible.}

\newcommand\submittednotice{%
\begin{tikzpicture}[remember picture,overlay]
\node[anchor=south,yshift=10pt] at (current page.south) {\fbox{\parbox{\dimexpr0.65\textwidth-\fboxsep-\fboxrule\relax}{\submittedtext}}};
\end{tikzpicture}%
}

\begin{document}

\title{Ultrabroadband Gain-Switched and Superluminescent Terahertz Semiconductor Lasers}

\author{Urban Senica, Michael A. Schreiber, Mattias Beck, \\
Christian Jirauschek, J\'er\^ome Faist, \textit{Fellow, IEEE}, and Giacomo Scalari
\thanks{
\textit{Corresponding authors: Urban Senica, Giacomo Scalari.}

Urban Senica, Mattias Beck, J\'er\^ome Faist, and Giacomo Scalari are with the Institute for Quantum Electronics, ETH Z{\"u}rich, 8093 Z{\"u}rich, Switzerland. 

Urban Senica is also with the John A. Paulson School of Engineering and Applied Sciences, Harvard University, 02138 Cambridge, USA.

Michael A. Schreiber and Christian Jirauschek are with the TUM School of Computation, Information and Technology, Technical University of Munich (TUM), 85748 Garching, Germany.}}



\maketitle

\begin{abstract}
Terahertz quantum cascade lasers (THz QCLs) are chip-scale semiconductor lasers operating in the frequency range between 1-6 THz, useful as compact sources for spectroscopy, communications, and non-destructive imaging and testing. Here, we apply low-frequency microwave modulation on a planarized THz QCL to generate ultrabroadband emission in the THz range. For very low modulation frequencies below 1~GHz, a gain-switched octave-spanning spectrum with a smooth spectral envelope is generated between 1.9 - 4.1~THz. Increasing the modulation frequency broadens the lasing modes until a low-coherence, continuous emission spectrum is achieved in the superluminescent regime, covering the spectral region between around 3 - 4 THz, without any discrete lasing modes or spectral gaps. We complement the experimental results with extensive analytical models and numerical simulations that capture the intracavity laser dynamics and fully explain the different operation regimes. These devices could prove useful for absorption spectroscopy without any spectral gaps, and as ultrabroadband sources of THz radiation.
\end{abstract}


\section{Introduction}
\IEEEPARstart{T}{erahertz} quantum cascade lasers (THz QCLs)~\cite{kohler_terahertz_2002,williams_terahertz_2007} are chip-scale semiconductor lasers that can generate emission between 1.2-6 THz, spanning from single-mode~\cite{amanti2007, bosco2016} to frequency comb operation~\cite{burghoff_terahertz_2014, faist2016quantum} and ultrabroadband, octave-spanning emission~\cite{rosch_octave-spanning_2015,RoeschNanophotonics2018}. In recent years, considerable progress has also been made in the area of ring lasers~\cite{jaidl2021comb} and solitons~\cite{micheletti2023terahertz}, frequency-modulated combs~\cite{senica2023frequency, roy2024fundamental}, mode-locked pulses~\cite{wang2017short,senica2026continuously}, surface-emitting devices~\cite{curwen_terahertz_2018, curwen2019broadband, senica2023broadband}, and high-temperature operation~\cite{bosco_thermoelectrically_2019, khalatpour2021high, khalatpour2023enhanced,GloorHHLcoolerNanophot2025}.

The key performance aspects of THz sources in general are the generation of (ultra)broadband emission with smooth spectral envelopes and a sufficient spectral resolution. The latter is typically defined by the physical cavity length and the inversely proportional repetition rate that determines the spectral mode spacing. In this work, we overcome these practical limitations by injecting microwaves into the laser cavity to facilitate a strong modulation of the laser bias voltage at low frequencies, below the natural cavity repetition rate $f_\mathrm{rep}$. 

We use planarized THz~QCLs~\cite{senica2022planarized}, where the active region waveguide is embedded within the low-loss polymer benzocyclobutene (BCB) and covered by an extended top metallization, which enables the placement of bonding wires over the passive area, as shown in Fig. \ref{fig:SlowModulation_Gaussian}(a). Since the laser cavity is essentially embedded within a microwave waveguide, this makes it especially suitable for low-loss microwave modulation experiments, where a radio frequency (RF) source and amplifier are used to modulate the laser bias voltage via a set of short bonding wires at the back waveguide facet. The laser sample is mounted on a Helium flow cryostat cold finger and measured in a vacuum at cryogenic heat sink temperatures, typically between 20 - 40~K.  The emission spectrum is acquired using a Fourier transform infrared (FTIR) spectrometer, which contains a Michelson interferometer and a room-temperature deuterated triglycine sulfate (DTGS) detector. The resulting spectral interferogram displays a pattern of bursts, with a periodicity inversely proportional to the laser repetition frequency $f_\mathrm{rep}$. Its shape and amplitude already reveal the basic information about the laser state, e.g., a coherent comb state will necessarily produce symmetric, non-decaying interferograms \cite{forrer2021self}, while a non-symmetric and/or decaying interferogram indicates an incoherent state. With a Fourier transform of an interferogram, we can reconstruct the measured THz emission spectrum.

In the following sections, we first examine devices fabricated from a broadband homogeneous active region \cite{forrer_photon-driven_2020} to present the initial results and the analytical and numerical models that explain the underlying physical phenomena. Additionally, in the final results section, we use a broadband heterogeneous active region \cite{RoeschNanophotonics2018} in combination with a modified microwave waveguide and high-power RF amplifiers to produce the best-performing devices.

\submittednotice

\section{Gain switching}
\subsection{Experimental results} 
When applying microwave modulation at very low frequencies, significantly below the natural repetition rate (typically at around $f_\mathrm{mod}\approx$ 1 GHz), a broadband emission spectrum is generated. It can span the entire laser gain bandwidth of the active region with a smooth spectral envelope, where the lasing modes are separated by the usual free-running repetition rate $f_\mathrm{{rep}}=\frac{c}{2\,n_\mathrm{g}\,L}$, where $c$ is the speed of light in vacuum, $n_\mathrm{g}$ is the group refractive index of the propagating optical mode, and $L$ is the cavity length.

\begin{figure}[tb]
\centering
\includegraphics[width=1\linewidth]{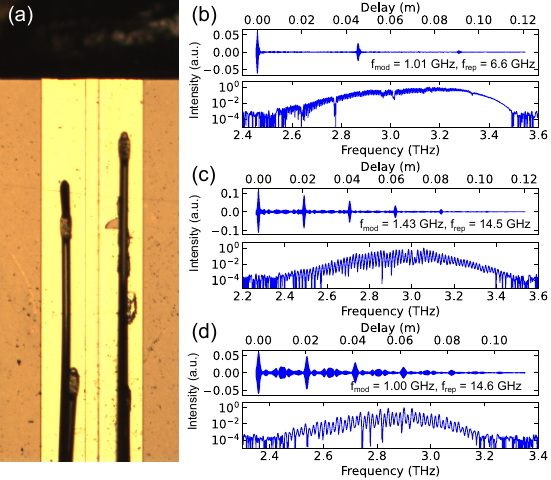}
\caption[Slow modulation for broadband emission spectra]{\textbf{(a)} Top view optical microscope image of a mounted planarized terahertz quantum cascade laser (THz QCL). The central narrow stripe is the active waveguide, with bonding wires attached to the extended top metallization over the BCB polymer. The THz emission is collected and measured from the front cleaved facet. When a slow microwave modulation ($f_\mathrm{{rep}}/20\lesssim f_\mathrm{{mod}}\lesssim f_\mathrm{{rep}}/7$) is applied to a device biased close to the lasing threshold, a broadband spectrum is generated that can span the full gain bandwidth and has a smooth spectral envelope. Here, we show results on three different devices: \textbf{(b)} a long homogeneous ridge waveguide, \textbf{(c)} a dispersion-compensated ridge waveguide, and \textbf{(d)} an asymmetric Y-coupled device.}
\label{fig:SlowModulation_Gaussian}
\end{figure}

In terms of microwaves, we are in a quasi-static regime, as the period of the driving microwave signal is usually in the range of 7 - 20 round-trip times, resulting in a uniform microwave field distribution across the entire laser cavity. This is a slow modulation of the laser bias, resulting in switching the laser periodically on and off for several round-trip times. 
In Fig. \ref{fig:SlowModulation_Gaussian}(b-d), we show the decaying interferograms (a sign that the state is incoherent/periodically switching off), and the corresponding broadband emission spectra of three different devices (planarized ridge waveguide~\cite{senica2022planarized}, dispersion-compensated device, and an asymmetric Y-coupled device~\cite{senica2026short}). All of these devices are in a gain-switched regime~\cite{lau1988gain,jukam2009terahertz}, and the emission is incoherent as the laser is switched off for several round-trip times between subsequent pulses, which are fully absorbed before the start of the next pulse. Similar spectra were recently also observed in mid-IR QCLs  both with a slow microwave modulation~\cite{CargioliAPLPhot2024} and with rapidly gain-switched short pulses  \cite{taschler2023short}.

This approach can be used to reliably generate broadband spectra with smooth envelopes that span the entire gain bandwidth, for any kind of planarized waveguide sample. This is usually not the case in free-running devices, where only a constant DC bias voltage and current are applied, which often display narrower bandwidths and spectral envelopes with varying magnitudes.
To provide an intuitive picture of the physical mechanism, one can think of a laser that is turned on for such a brief period of time that the nonlinear effects (e.g., spatial hole burning, mode competition, and four-wave mixing) that dominate at longer timescales are not as pronounced \cite{CargioliAPLPhot2024}. In the case of turning the laser on for extended periods of time (several thousand round trips or even continous-wave operation), the mode competition and intrinsic nonlinearities will self-stabilize the laser in a highly coherent frequency comb state, but with a limited emission bandwidth. In our regime of low-frequency microwave modulation, the laser state has a lower temporal coherence but features much greater emission bandwidths.

\subsection{Simulation Results} In order to gain additional insight into the experimental observations, we perform simulations based on a semiclassical Maxwell-Bloch approach~\cite{allen1987optical,jirauschek2019optoelectronic}. Therein, the optical field is modeled using 1D Maxwell's equations, where the rotating wave approximation (RWA) is applied to reduce the computational load. To this end, we make an ansatz for the 1D electric field 

\begin{equation}
\begin{aligned}
    E(z,t) &= \mathcal{R} \big\{ E^{+}(z,t) \,  \mathrm{e}^{\mathrm{i} ( k_\mathrm{c} z - \omega_\mathrm{c} t ) } \\ 
    & \phantom{=}+ E^{-}(z,t) \,  \mathrm{e}^{\mathrm{i} (-k_\mathrm{c} z -\omega_\mathrm{c} t)} \big\} \, , 
    \label{eq:RWA_ansatz}
\end{aligned}
\end{equation}
\noindent
where $E^{\pm}$ are the complex field envelopes of the forward~$(+)$ and backward~$(-)$ traveling waves. Moreover, $\omega_\mathrm{c}$ denotes the center frequency and $k_\mathrm{c} = \omega_\mathrm{c} n / c$ is the corresponding propagation constant, with the background refractive index $n$. With this, we arrive at the propagation equation for the field envelopes~\cite{jirauschek2019optoelectronic}

\begin{equation}
\begin{aligned}
    \partial_t E^\pm \pm v_\mathrm{g} \, \partial_z E^\pm = & -l E^\pm + f^\pm \\
    &- \frac{\mathrm{i}v_\mathrm{g}}{2} \, \beta_2 \, \partial_t^2 E^\pm + S^\pm \, .
    \label{eq:optical_propag}
\end{aligned}
\end{equation}

Here, $v_\mathrm{g}$ is the group velocity, and $l =  v_\mathrm{g} a_\mathrm{w} / 2$, where $a_\mathrm{w}$ represents the waveguide losses. Additionally, our model incorporates the background group velocity dispersion $\beta_2$ and the polarization $f^\pm$ induced by the QCL's active region. As spontaneous emission is not naturally included in a semiclassical description~\cite{stowasser2024stochastic}, we introduce it by adding the noise terms $S^\pm$ to the field propagation equation. These noise terms are implemented by adding random numbers to the numerical field update equations, whose power spectrum reproduces a Lorentzian lineshape function~\cite{jirauschek2023dynamicSpont}. 

The laser's active region is modeled as a two-level quantum system using a density-matrix approach. By choosing a representation of the density matrix as in Eq.~(\ref{eq:RWA_ansatz}) and applying the RWA, one obtains the standard Bloch~equations~\cite{jirauschek2019optoelectronic}

\begin{equation}
\begin{aligned}
    \partial_t \eta_{21}^\pm &= \mathrm{i} \Delta \eta_{21}^\pm - \frac{\mathrm{i} d_{21}}{2 \hbar} \left( w_0 \, E^\pm  + w^\pm \, E^\mp \right) - \gamma_2 \, \eta_{21}^\pm \, ,\\
    \partial_t w_{0} &= \frac{2}{\hbar} \,  \mathcal{I} \left\{ d_{21}^* \, [\eta_{21}^+ (E^+)^* + \eta_{21}^- (E^-)^* ] \right\} \\
    & \phantom{=} -\gamma_1 ( w_0 - w_\mathrm{eq} ) \, , \\
    \partial_t w^+ &= \frac{\mathrm{i}}{\hbar} [d_{21} (\eta_{21}^-)^* \, E^+ - d_{21}^* \eta_{21}^+ \, (E^-)^* ] \\
    & \phantom{=} - \gamma_1 w^+ - 4 k_\mathrm{c}^2 D w^+ \, ,
\end{aligned}
\end{equation}
\noindent
for the coherence terms $\eta_{21}^\pm(z,t)$ and the population inversion $w(z,t) = w_0(z,t) + 2 \mathcal{R} \left\{ w^+(z,t) \, \mathrm{exp}(2\mathrm{i} k_\mathrm{c} z) \right\}$, which consists of the average inversion $w_0$ and the inversion $w^+=(w^-)^*$ associated with spatial hole burning. Furthermore, $\hbar$, $\Delta$, $d_{21}$, $D$, $w_\mathrm{eq}$ and $\gamma_1$, $\gamma_2$ denote the reduced Planck's constant, the detuning $\Delta = \omega_\mathrm{c} - \omega_{21}$ between $\omega_\mathrm{c}$ and the resonance frequency of the optical transition $\omega_{21}$, the dipole matrix element, the diffusion constant, the equilibrium inversion, and the gain recovery and dephasing rates, respectively. The polarization from Eq.~(\ref{eq:optical_propag}) is then directly obtained from the Bloch equations as $f^\pm \propto \eta_{21}^\pm$~\cite{jirauschek2019optoelectronic}.

An injected modulation signal at the back side of the device ($z=0$) leads to a microwave propagating along the waveguide contacts. We include this effect by analytically solving the transmission line equations for a Fabry–Pérot cavity, and obtain the solution for the microwave field as

\begin{equation}
     \Delta u(z,t) = U_\mathrm{mod} \, \mathcal{R} \left\{ \frac{ \cos \left[ \beta_\mathrm{mod} \, (z-L) \right] }{\cos{ (\beta_\mathrm{mod} \, L) }} \,  \mathrm{e}^{-\mathrm{i}\omega_\mathrm{mod} t} \right\} \, ,
\end{equation}
\noindent 
where $\omega_\mathrm{mod} = 2 \pi f_\mathrm{mod}$ denotes the modulation frequency, $U_\mathrm{mod}$ represents the amplitude of the modulation field, and $\beta_\mathrm{mod}$ corresponds to the complex propagation constant (for more details, we refer to Ref.~\cite{schreiber2025dynamic}).

\begin{figure}[tb]
\centering
\includegraphics[width=0.8\linewidth]{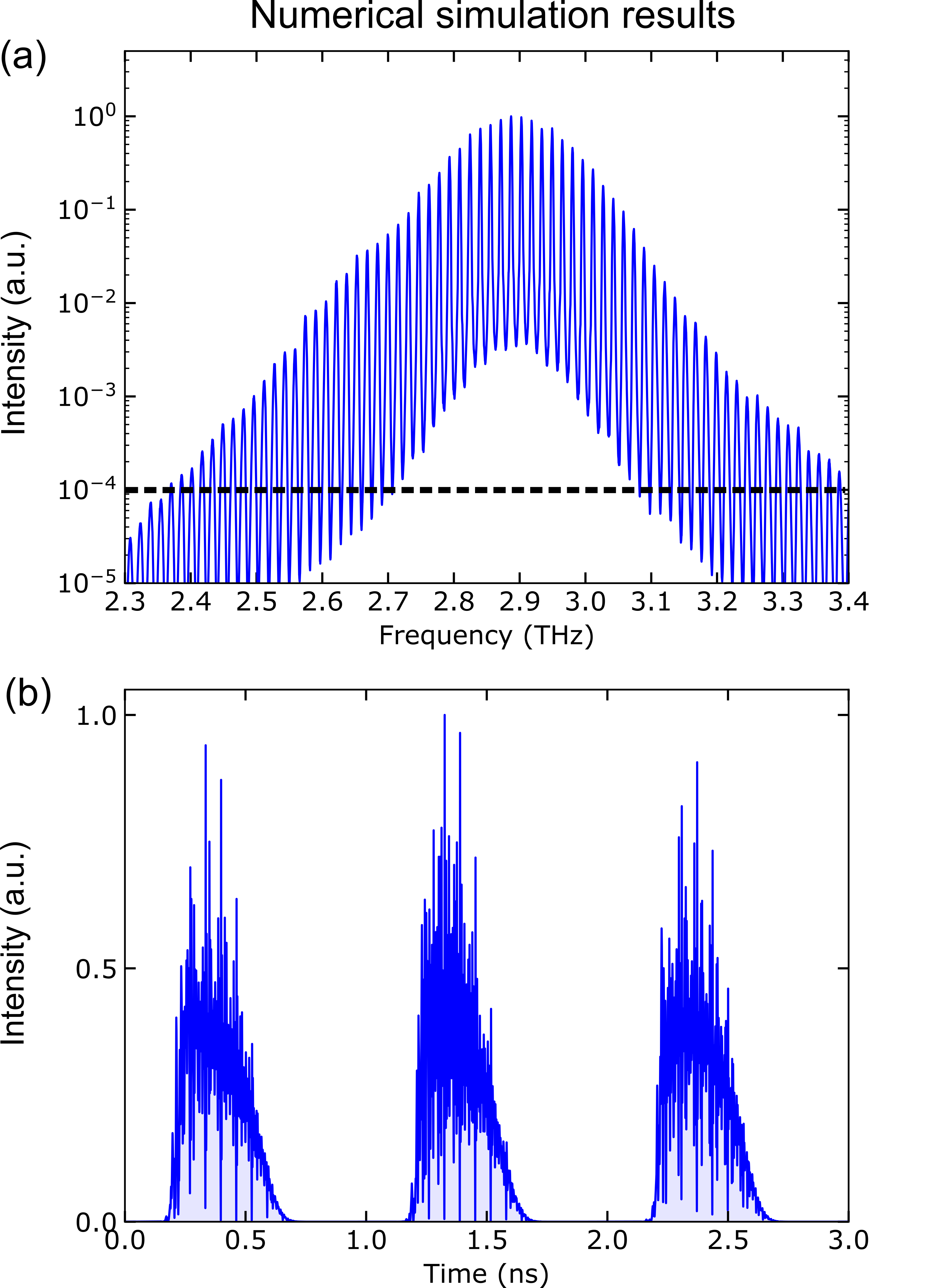}
\caption{Numerical simulation results of a gain-switched THz QCL: \textbf{(a)} displays the simulated spectrum, where a smoothing filter was applied to account for the spectral resolution of the experimental measurement. The black dashed line indicates the experimental signal-to-noise ratio of around 40 dB. \textbf{(b)} shows the corresponding time-domain data. Here, a train of separated incoherent pulses with strong intensity fluctuations is visible, as each pulse is amplified from spontaneous emission, with no correlation to the previous pulse.}
\label{fig:gainswitch_sim}
\end{figure}

In Fig.~\ref{fig:gainswitch_sim}, we present simulation results for a $L~=~2.7$~mm long THz QCL, which is modulated at a frequency $f_\mathrm{mod} = 1.0$~GHz $\ll f_\mathrm{rep}$. To compare our simulated spectrum (see Fig.~\ref{fig:gainswitch_sim}~(a)) with the experimental findings, we apply a smoothing filter to the data in order to account for the limited spectral resolution of the measurement setup. This results in a spectrum with broadened modes separated by the natural repetition rate of the device $f_\mathrm{rep}$. The bandwidth and the shape of our simulated spectra are in good agreement with the experimental data from Fig.~\ref{fig:SlowModulation_Gaussian}~(d).

Our numerical simulations also allow us to reveal the time-domain waveforms of the gain-switched device, which are displayed in Fig.~\ref{fig:gainswitch_sim}~(b). Clearly separated pulses with strong amplitude fluctuations are visible, as is expected from an incoherent gain switching regime~\cite{taschler2023short}. We note that the on- and off-times ($T_\mathrm{on}$ and $T_\mathrm{off}$) of the pulses are almost equally long, and correspond to half the period length of the modulation signal, i.e., $ T_\mathrm{on} \approx T_\mathrm{off}\approx 1/ (2f_\mathrm{mod}) \approx 0.5$ ns. Hence, the laser is turned on for several cavity round-trip times $1/f_\mathrm{rep}= 2 n_\mathrm{g} L / c$, and subsequently switched off again for the same amount of time. Consequently, the optical intensity within the laser cavity completely decays and has to build up again from spontaneous emission noise (incorporated in Eq.~(\ref{eq:optical_propag})), resulting in an incoherent lasing state. The intensity fluctuations on each pulse can approximately be associated with the round-trip time of the laser cavity $1/f_\mathrm{rep}$, leading to the aforementioned mode spacing of $f_\mathrm{rep}$ in the frequency domain.

\section{Superluminescence}
\subsection{Experimental results}
\begin{figure}[tb]
\centering
\includegraphics[width=0.9\linewidth]{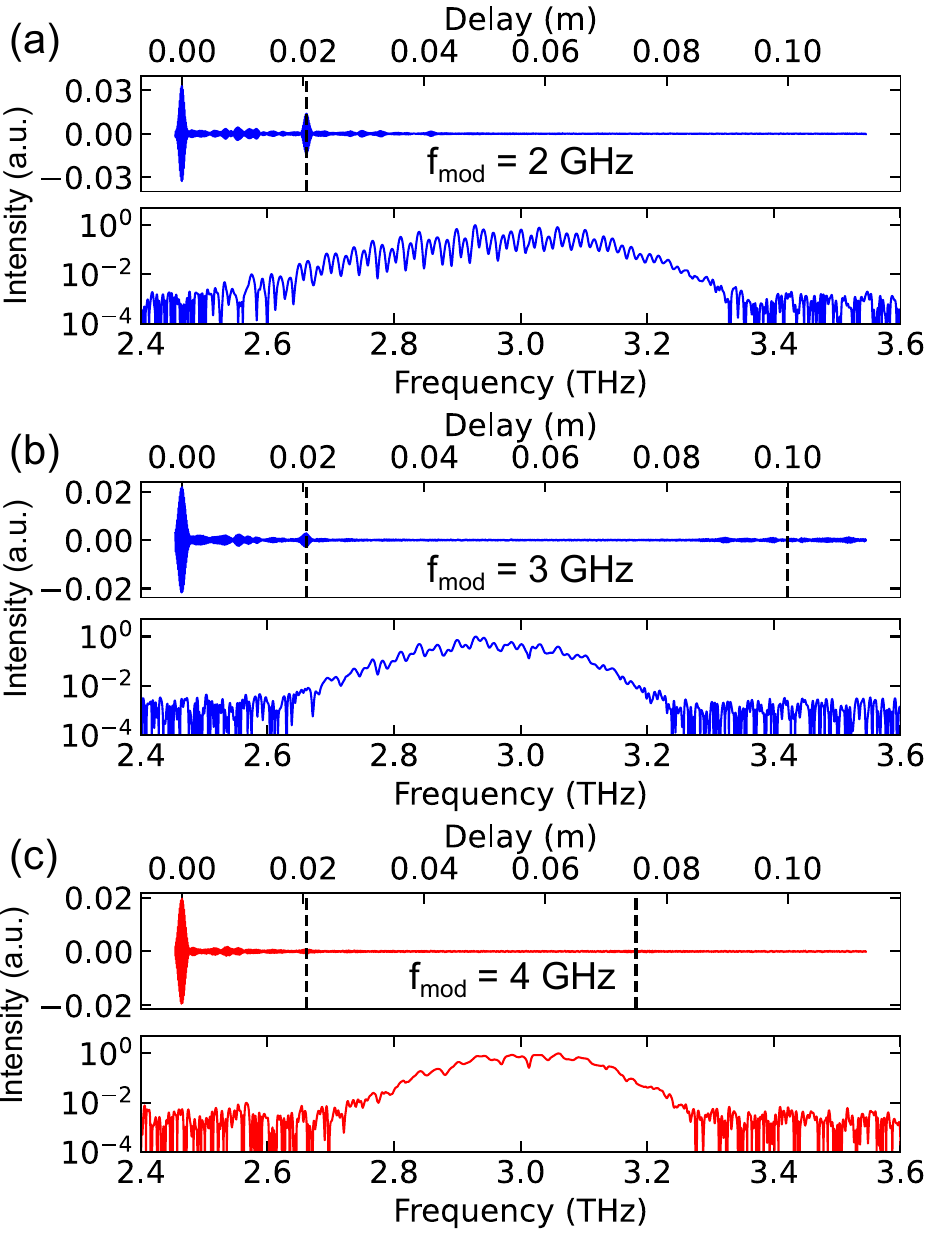}
\caption[Amplified spontaneous emission with slow modulation]{By increasing the modulation frequency of a sample biased close to the threshold, the lasing modes are broadened and eventually merge to a continuous spectrum without any discrete lasing modes visible. In the interferograms, the first vertical dashed line marks the expected position of the second burst for modes separated by $f_\mathrm{{rep}}$, while the second line indicates the expected position of a third burst for modes separated by $f_\mathrm{{mod}}$. In panel (c), there is no intensity at this third burst, confirming the generation of a continuous spectrum with no discrete modes, i.e., we are not limited by the FTIR spectral resolution of 2.25 GHz, which would be enough to reveal a signal repetition frequency of 4 GHz.}
\label{fig:SlowModulation_supercontinuum}
\end{figure}

In Fig. \ref{fig:SlowModulation_supercontinuum}, we display the interferograms and emission spectra of the device from Fig. \ref{fig:SlowModulation_Gaussian}(d) for an increasing modulation frequency (the free-running repetition rate is around $f_\mathrm{{rep}}\approx14.6$ GHz). As the modulation frequency is increased, the interferogram decays more rapidly, and the first burst at the zero path delay (ZPD) has a lower amplitude. In the spectrum, the width of the reconstructed modes is broadened. Only the first burst at ZPD remains for an injection frequency of $f_\mathrm{{mod}}=4$ GHz. The corresponding spectrum is continuous, without any discrete modes visible. We should emphasize that we can rule out the trivial case of observing a continuous spectral envelope due to a limited spectral resolution of our spectrometer: if discrete lasing modes with a spacing of the injected RF signal of 4 GHz were present as a result of a sideband generation process, this would already fall within the spectral resolution of the measurement of 2.25 GHz: the second dashed vertical line indicates the expected position of the additional burst. 

The device is in a superluminescent (amplified spontaneous emission) regime, where the emission is incoherent, producing a broadband and continuous emission spectrum. In superluminescent diodes \cite{alphonse1988high}, this is often achieved by using a laser waveguide with very low cavity feedback (e.g., with antireflection coatings and tilted end facets), such that there is a large single-pass gain for the spontaneous emission, but lasing action is suppressed due to high mirror losses.
In our case, we do not use low-reflectivity facets to suppress lasing, but an RF modulation, where the frequency is around $f_\mathrm{{mod}}\approx f_\mathrm{{rep}}/3.7$. Since the DC bias voltage is at the laser threshold, the RF modulation only induces a positive net gain for the positive half of the modulation, which is less than two round-trip times. 

\subsection{Analytical model: amplifier with gain modulation}
In order to reproduce the results from Fig. \ref{fig:SlowModulation_supercontinuum}(c), we developed a simple model of an amplifier with gain modulation defined by the RF injection, where several simplifying assumptions were made. First of all, we consider a uniform RF field distribution across the cavity, that the gain is proportional to the RF-modulated laser bias (with an instantaneous response), and that there is no gain saturation. Additionally, we assume a constant (frequency-independent) refractive index $n$.

\begin{figure}[tb]
\centering
\includegraphics[width=1\linewidth]{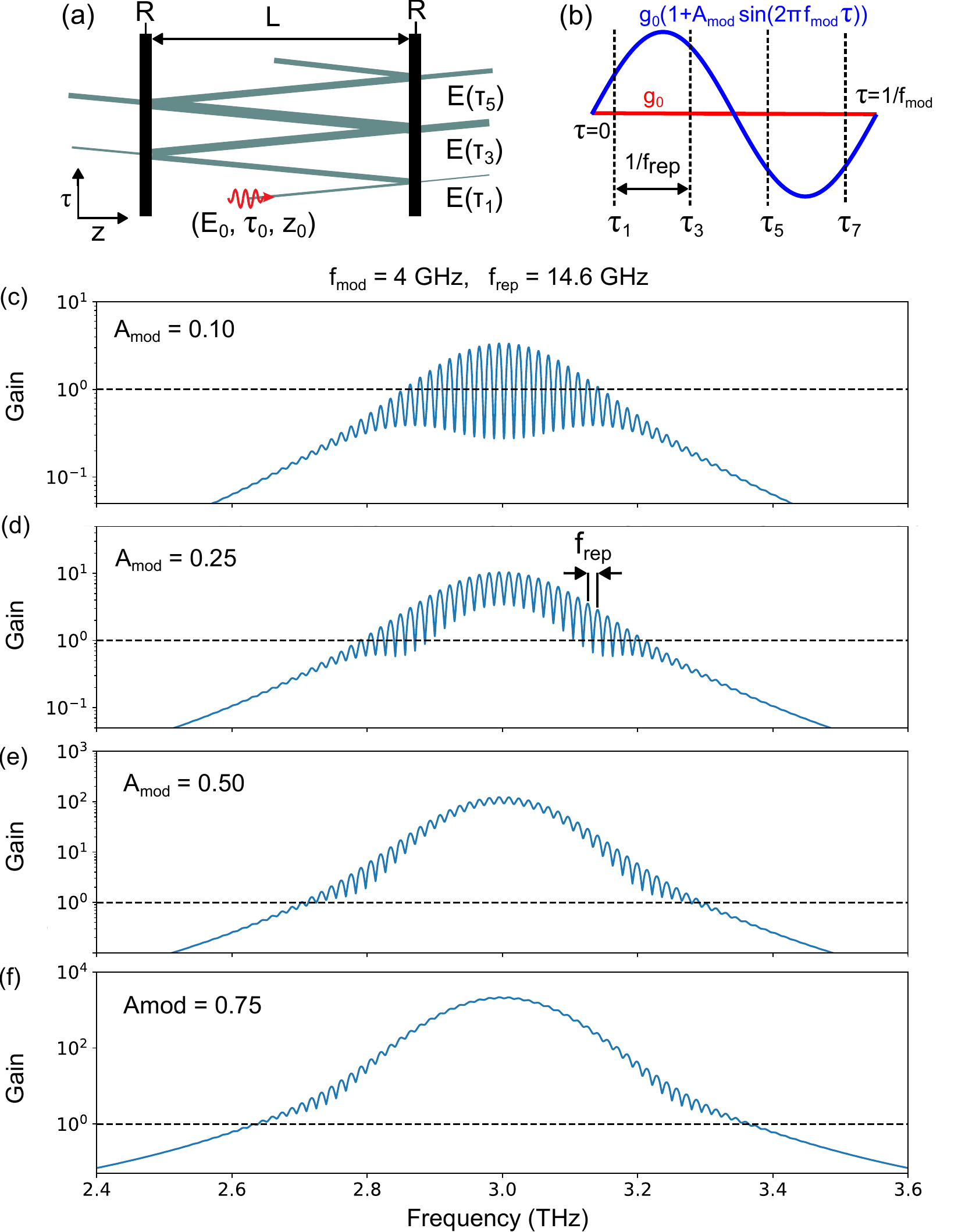}
\caption[Simulated frequency-dependent gain of an amplifier with gain modulation]{Schematics and results of the model of an amplifier with gain modulation. \textbf{(a)} Cavity model: spontaneous emission is propagating between the partially-reflective mirrors, with a frequency-dependent accumulated phase and a frequency- and time-dependent amplitude. The output field is an interference between waves separated by a round-trip time. \textbf{(b)} The gain is modeled to follow the RF modulation at frequency $f_\mathrm{{mod}}$ and amplitude $A_\mathrm{{mod}}$, while the laser is biased at the threshold, $g_0=\alpha_\mathrm{{wg}}+\alpha_\mathrm{{m}}$. Due to the time-dependent gain, the amplitudes of the interfering waves at $\tau_1, \tau_3, \tau_5$ and $\tau_7$ are computed with an integral. Below, we plot the simulated frequency-dependent gain of an amplifier with gain modulation, for the same parameters as in the measurements (the black dashed line marks the region with unity intensity gain). For an increasing RF-modulation amplitude, the mode contrast is reduced and the intensity gain is increased. The results in panel (e) are in good agreement with the measurements in Fig. \ref{fig:SlowModulation_supercontinuum}(c), including the emission spectrum shape and bandwidth, and the 20 dB amplification at the center of the spectrum.}
\label{fig:ASE_sim_gain}
\end{figure}

In Fig. \ref{fig:ASE_sim_gain}(a), we show a schematic illustration of the modeled system: a Fabry-Pérot cavity with a length $L$ and finite mirror reflectivities $R$. Spontaneous emission that couples into a cavity mode will propagate and bounce between partially reflective end mirrors. The output field will be determined by the interference of the fields separated by one full round-trip time $1/f_\mathrm{{rep}}=\frac{2L n_\mathrm{g}}{c}$. In a passive Fabry-Pérot cavity (no gain or loss), an input optical wave will generate sharp intensity maxima and minima as a function of frequency due to constructive and destructive interference. In our case, there is a time-dependent gain/loss, so the amplitudes of the interfering fields will vary between round-trip times. We can calculate the field amplitude gain of spontaneous emission generated at position $z_0$, time $\tau_0$ and frequency $f$ propagating towards the front mirror, by computing the interference between the output fields separated by the round-trip time:

{\small
\begin{align}
&\mathrm{G_\mathrm{E}}(f,z_0) = \frac{E_\mathrm{out}}{E_0} \nonumber \\ 
&= \underbrace{\exp\bigg\{\int_0^{\tau_1} (g(f,\tau)-\alpha_\mathrm{{wg}}) \frac{c}{n} \mathrm{d}\tau\bigg\}}_{\text{amplitude}}\ \underbrace{\exp\bigg\{i  \frac{2\pi f n}{c}(L-z_0)\bigg\}}_{\text{phase}} \times 
    \nonumber \\
    &\bigg(\underbrace{t}_{\tau_1}+\underbrace{r^2 t\ \exp\!\bigg\{\int_{\tau_1}^{\tau_3} (g(f,\tau)\!-\!\alpha_\mathrm{{wg}}) \frac{c}{n} \mathrm{d}\tau\bigg\}\!\exp\!\bigg\{i\frac{2\pi f n}{c}2L\bigg\}}_{\tau_3} \nonumber \\
    &+\underbrace{\ldots }_{\tau_N}\bigg)
\end{align}
}%
where $t, r$ are the complex field transmission and reflection coefficients, $\tau_3-\tau_1$ is the round-trip time, and $g(f,\tau)$ is the modulated gain:
\begin{align}
    g(f,\tau) = g_0(f)\ (1+A_\mathrm{{mod}} \sin(2\pi f_\mathrm{{mod}}\tau))
\end{align}
The sum is performed for a single RF modulation cycle, so the number of single passes in the calculation is limited to $m = \lfloor(f_\mathrm{{rep}}/f_\mathrm{{mod}})\rfloor$.
The amplitude gain at frequency $f$ for all the starting points $z_0$ is obtained with $\frac{1}{M}\sum_{z_0=0}^L G_\mathrm{E}(f,z_0)$, where $M$ is the number of evaluated points $z_0$ (in order to get a normalized gain value). This whole calculation is also done for an initial propagation in the other direction (towards the back mirror), where we sum up all the even transmissions at times $\tau_2, \tau_4$, etc. Finally, to obtain the normalized frequency-dependent gain curve, we sweep the input frequency $f$. Since in the experiments, we measure the output field intensity, we can calculate the intensity gain via the relation $G_\mathrm{I} = |G_\mathrm{E}|^2$.

We then computed the frequency-dependent intensity gain curves for our specific device with a length of $L=2.7$ mm, a repetition rate $f_\mathrm{{rep}}=14.6$ GHz, and an RF modulation frequency of 4 GHz. In this case, we need to interfere the fields of four consecutive round-trip times, as sketched in Fig. \ref{fig:ASE_sim_gain}(b). The frequency-dependent active region (material) gain was defined as a Lorentzian gain curve centered at 3 THz with a FWHM width of 1 THz, where the peak of the gain was set to match the losses ($\alpha_\mathrm{{wg}}+\alpha_\mathrm{{m}}$), as the sample is biased at the laser threshold. In Fig. \ref{fig:ASE_sim_gain}(c-f), we show the results: for an increasing RF-modulation amplitude, the contrast between the modes is reduced and the gain is increased. For very large modulation amplitudes, a continuous spectrum is obtained, in good agreement with the measurement results in Fig. \ref{fig:SlowModulation_supercontinuum}(c). 

The explanation is relatively straightforward: for an increasing modulation amplitude, the difference between the amplitudes of interfering fields increases, up to a point where their relative phases do not play a role anymore. In other words, a continuous/smooth intensity gain is obtained for any frequency, as the cavity no longer acts as a frequency-selective filter.
During the negative half-cycle of the RF modulation (nearly two round-trip times), there is a net loss and the circulating light is absorbed. The next positive half-cycle of the RF modulation then amplifies new spontaneous light - in the time domain, we get a sequence of incoherent pulses. The simulation result from Fig. \ref{fig:ASE_sim_gain}(e) is in good agreement with the measured 20 dB dynamic range and the bandwidth of the emission spectrum in Fig. \ref{fig:SlowModulation_supercontinuum}(c), suggesting that it is indeed an amplification within a single RF-modulation cycle.

Although incoherent, these kinds of states could be useful as broadband ``white light'' THz sources without any spectral holes or gaps. They could complement photoconductive antennas at higher THz frequencies, as the obtainable output powers of the latter typically drop rapidly above 1 THz \cite{burford2017review}. 

\begin{figure}[tb]
\centering
\includegraphics[width=0.9\linewidth]{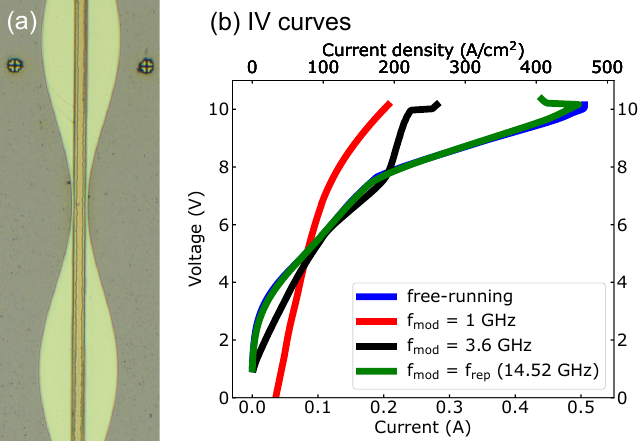}
\caption{\textbf{(a)} Optical microscope image of a planarized THz QCL with a sinusoidally-modulated extended top metallization, combining low-loss microwave propagation with local microwave field enhancement. \textbf{(b)} Measured IV curves of the device, displaying a significant dependence on the modulation frequency.}
\label{fig:sine_IV}
\end{figure}

\begin{figure*}[tb]
\centering
\includegraphics[width=0.9\linewidth]{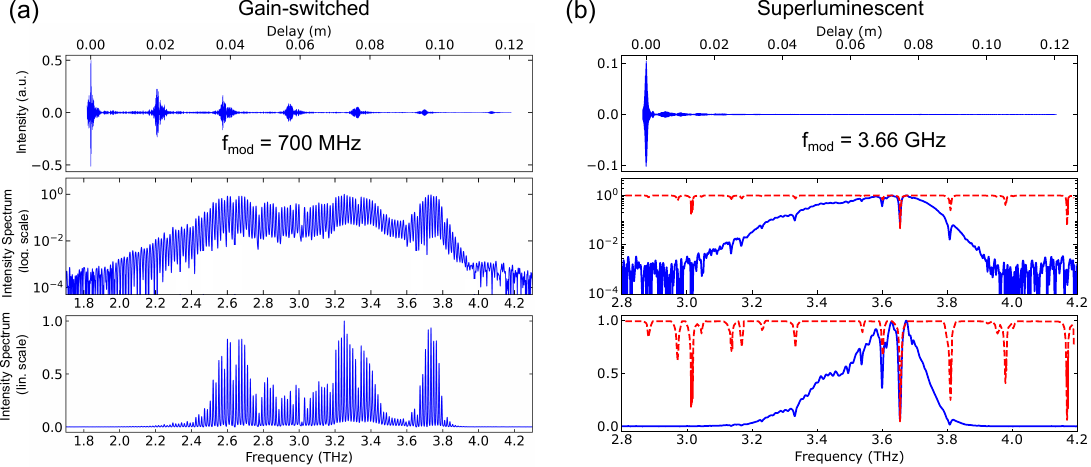}
\caption[Simulated frequency-dependent gain of an amplifier with gain modulation]{\textbf{(a)} Ultrabroadband gain-switched device operation with a modulation frequency of $f_\mathrm{mod}=700$ MHz. The panels show the interferogram and the emission spectrum in logarithmic and linear scales, respectively. The octave-spanning emission spectrum features a smooth spectral envelope across the entire range from 1.9 to 4.1 THz. \textbf{(b)} Superluminescent operation is achieved with a modulation frequency of $f_\mathrm{mod}=3.66$ GHz. The resulting continuous emission spectrum spans nearly 1 THz, with visible absorption lines due to parasitic absorption in the optical beam path. The absorption dips show good agreement with the atmospheric absorption lines (red).}
\label{fig:ultrabroadband}
\end{figure*}

\section{Ultrabroadband emission}
To explore the limits of these two regimes, we fabricated another planarized waveguide sample using the broadband heterogeneous active region~\cite{RoeschNanophotonics2018}. Additionally, we introduce a novel microwave waveguide design: by modifying the top extended metallization to follow a sinusoidally-modulated shape, we combine the low-loss microwave propagation of planarized waveguides with local microwave field enhancement. The optical microscope image of such a device is shown in Fig. \ref{fig:sine_IV}(a). Additionally, we employ high-power RF amplifiers, which deliver around +40 dBm of RF power at the amplifier output. All of these combined effects have a significant influence on the laser operation. We studied this by measuring the laser IV curves at different operating conditions. As shown in Fig. \ref{fig:sine_IV}(b), resonant modulation at $f_\mathrm{mod}=f_\mathrm{rep}$ results in a very similar dependence as a free-running device, with a slightly reduced threshold current (kink in the IV curve at around 190 mA, 7.8 V). In contrast, low-frequency modulation at 3.6 GHz (superluminescence) and especially at 1 GHz and below (gain switching) significantly alters the IV curve, increasing its slope. Remarkably, in this device and configuration with extreme microwave modulation, the broadband gain-switched and superluminescent regimes can be accessed for virtually any DC laser bias, not limited to a range close to the laser threshold. In fact, gain-switched states were observed even when the laser was not connected to any DC bias source and only a strong microwave modulation below 1 GHz was applied.

The resulting emission spectra for this device reach considerably broader bandwidths than those presented in previous sections. In particular, in Fig. \ref{fig:ultrabroadband}(a), we display the results of a gain-switched device, obtained with a low modulation frequency of 700 MHz. The emission spectrum covers more than a full octave of bandwidth, between around 1.9 - 4.1 THz, with a smooth spectral envelope across the entire multi-lobed heterogeneous gain curve. The spectral amplitudes are relatively flat over a broad bandwidth, as showcased in the bottom panel, where we display the emission spectrum in a linear scale.

As shown in Fig. \ref{fig:ultrabroadband}(b), by modulating the device at a frequency of 3.66 GHz, we obtain a broad superluminescent spectrum, spanning nearly 1 THz between around 3-4 THz, with a considerable bandwidth even in the linear scale plot in the bottom panel. As the spectrum is continuous, we can observe clear absorption features, originating from parasitic absorption in the optical beam path between the device and the FTIR detector (air, water vapour, nitrogen, etc.). These have a good agreement with atmospheric absorption lines (dashed red line in the spectral plots)\cite{wu2023towards}. This result highlights the potential application for high-resolution absorption spectroscopy without any spectral gaps.

\section{Conclusion}
In conclusion, we demonstrated, both in experiment and simulation, how low-frequency microwave modulation of planarized THz QCLs can be employed to generate ultrabroadband emission spectra. In particular, very low modulation frequencies of $f_\mathrm{{rep}}/20\lesssim f_\mathrm{{mod}}\lesssim f_\mathrm{{rep}}/7$ (typically below 1 GHz) can drive the device into a gain-switched regime, where the laser is periodically turned on and off in the range of 7-20 round-trip times. In this case, we observe a very broad and smooth emission spectrum envelope that spans the entire active region gain bandwidth, with discrete lasing modes spaced at the natural cavity repetition rate $f_\mathrm{{rep}}$. In the time domain, this corresponds to incoherent pulses defined by the modulation frequency, with a typical duty cycle of around 50\% with sub-ns pulse durations.

Additionally, we discovered that increasing the modulation frequency gradually broadens the lasing modes, eventually reaching a superluminescent regime. In this case, we can generate broadband continuous emission spectra without any discrete modes, covering the entire bandwidth with no spectral gaps.

By combining a broadband heterogeneous active region with a field-enhancing sinusoidally-modulated top metallization and high-power RF amplifiers, we measured ultrabroadband octave-spanning gain-switched spectra and broadband superluminescent spectra spanning almost 1 THz. These devices will be useful for broadband absorption spectroscopy and as sources of white THz light. The superluminescent regime can be used to complement photoconductive antennas, since their spectral power density typically drops rapidly above 1 THz.

\section*{Acknowledgments}
U.S., G.S., M.B. and J.F. acknowledge the use of the FIRST-lab clean room facilities for laser growth and device fabrication, the SNF Project 200021-232335 for funding, as well as the EU Project iFLOWS. This work was partially performed within the framework of the 23FUN04 COMOMET project, supported by the European Partnership on Metrology, co-financed from the European Union’s Horizon Europe Research and Innovation Programme and by the Participating States. Funder ID 10.13039/100019599

\printbibliography


\end{document}